\documentclass[aps,prl,twocolumn,showpacs,superscriptaddress,groupedaddress]{revtex4-2}

\usepackage[utf8]{inputenc}

\usepackage[caption=false]{subfig}
\usepackage{graphicx} 
\usepackage{mathrsfs}
\usepackage{amsfonts}
\usepackage{amsmath}
\usepackage[noabbrev]{cleveref}
\usepackage[toc,page]{appendix}
\usepackage{import}
\usepackage[nice]{units}
\usepackage{float}
\usepackage{placeins}
\usepackage{color}

\begin{document}

\title{
Bridging cumulative and non-cumulative geometric frustration response via a frustrated $N$-state spin system.}
\date{\today}
\author{Snir Meiri}
\author{Efi Efrati} 
\email{efi.efrati@weizmann.ac.il} 
\affiliation{Department of
Physics of Complex Systems, Weizmann Institute of Science, Rehovot
76100, Israel}

\begin{abstract}
The resolution of geometric frustration in systems with continuous degrees of freedom often involves a cooperative inhomogeneous response and super-extensive energy scaling. In contrast, the frustration in frustrated Ising-like spin systems is resolved uniformly. In this work we bridge between these two extremes by studying a frustrated model composed of N-state spins, and varying N. The
expected cooperative response, observed for large N, is strongly attenuated as N is reduced, in a non-trivial way. Moderate N values show unique topological-like phases not observed before in frustrated models.
\end{abstract}
\maketitle
Geometric frustration arises whenever the short-range interactions between the constituents of a system favor the formation of local motifs that are incompatible with long range propagation in the ambient space geometry or topology. 
Examples of geometrically frustrated assemblies include: frustrated spin models \cite{Wan50,TJ83,Hal85,LBH+18} in which the frustration arises from a locally preferred misalignment of adjacent spins that cannot be reconciled over extended areas, and liquid crystals in which local tendencies of the constituents, such as a positive bend and vanishing splay and twist, necessitate the existence of unfavorable spatial structural gradients \cite{Mey76,NE18,HG20,Sel22}.
As the locally favored arrangement of the constituents cannot be realized, any finite assembly must exhibit some compromise of these locally favored tendencies. In some cases, such as the Ising antiferromanet on triangular lattice, this unavoidable compromise can be made uniform , i.e. distribute the frustration energy evenly by incorporating a single unfavorable bond in every facet. Consequently the energy associated with this compromise will grow extensively, proportional to the number of facets. Such frustration was recently termed non-cumulative \cite{ME21}, in contrast with systems exhibiting cumulative frustration, such as the frustrated XY-model \cite{ME22}. In the latter model the optimal compromise in a single triangular facet precludes repeating the same compromise verbatim in the adjacent facets. As a result, the energy associated with the compromise in systems exhibiting cumulative frustration grows super-extensively, and the ground-state shows long range cooperativity and depends on global properties of the system such as its aspect ratio and its spatial extent. The super-extensive energy scaling and its dependence on global parameters may lead to growth arrest \cite{TMH+21}, formation of filamentous structures \cite{MPNM14,LW17}, and give rise to exotic response properties \cite{SM21,LS16}. Naturally, the super-extensive energy scaling does not persist indefinitely and only applies to finite systems. 
Different types of geometric frustration may exhibit distinct mechanisms for frustration saturation for large systems, and in particular could result in infinite bulks of extensive energy, defect ridden structures or growth arrest along one or a few dimensions \cite{HG20}. 

Considering moderately sized frustrated systems \footnote{much larger than the scale of the constituents yet smaller than the typical geometric scale associated with the frustration, which in the present case reads $1/b_0$.}, in which the discrete DOF may assume a very large number of states, $N$, we expect to recover the continuous behavior that results in cumulative frustration. In contrast, as was recently shown, low $N$ values lead to a local and uniform resolution of the frustration, saturating at the level of a single or a few facets \cite{RL19}.
The present work aims to understand this discrepancy by filling in the gap between these two distinct and limiting behaviors. 
By varying the granularity of the model in a controllable way, we identify the threshold value of the number of DOF leading to the loss of uniformity and the incorporation of spatial gradients to the system and relate its value to the frustration. We show that as $N$ is increased the approach to the cumulative behavior is non-monotonous and report a previously undescribed topological-like phenomena at intermediate granularity.

\begin{figure*}[! t]
\includegraphics[width=16.0cm]{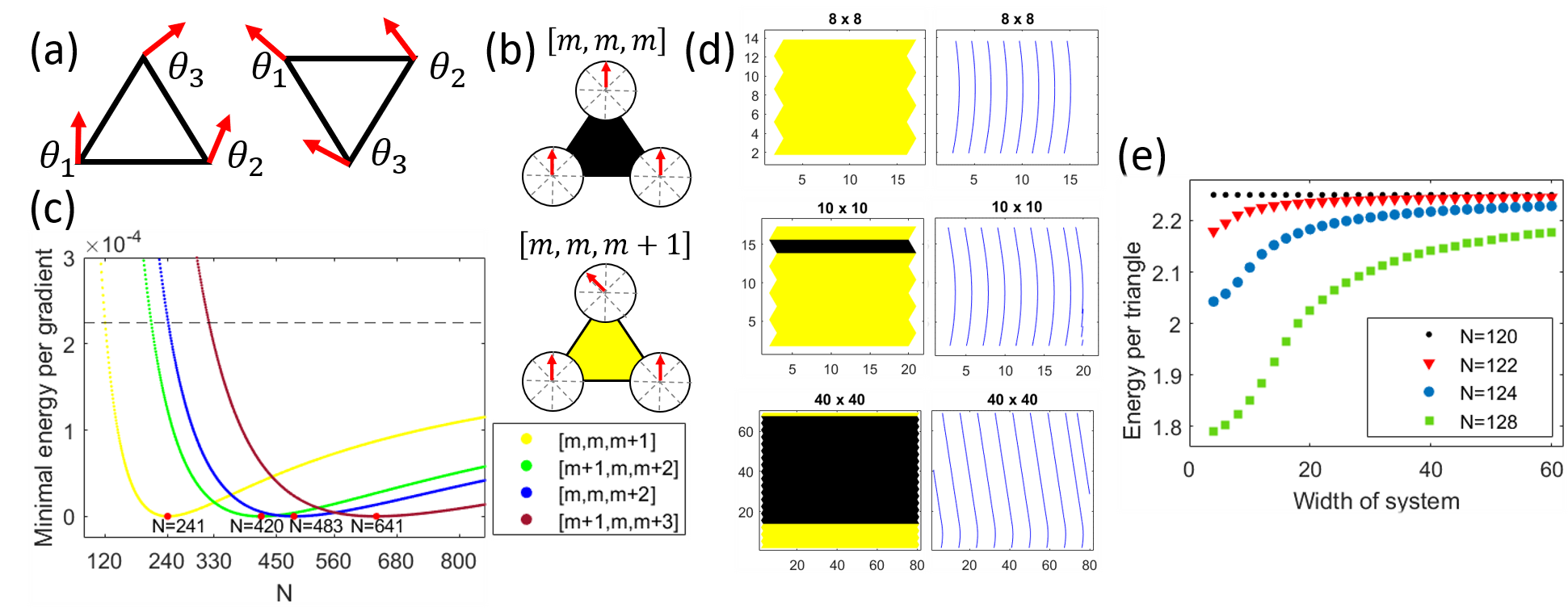}
\caption{ 
(a) Notations of vertices of upright (left) and up-side-down (right) triangles. (b) Color coding of two possible conformations of the spins. The state-key on top is defined such that each entry represents a state of a spin according to the notations in (a). $m$ is any integer between $1$ and $N$ where $N$ is the number of states in the model. (c) The minimal energy of a single triangle in a given conformation type for different $N$ values. The $N$ at which this minimal value is attained is marked. The dashed line marks the energy of the trivial constant solution. (d) Resulting minimal conformations for lattices of $8\times8$, $10\times10$ and $40\times40$ sites per edge in the case of $N=124$. Color coding of the left panels is according to the gradient types in (b) and the right panels display the interpolated streamlines. (e) Energy per triangle vs. the width of the system (number of sites per edge) for different values of $N$ near the cumulative-non-cumulative transition.}
\label{fig:model}
\end{figure*}

We consider a frustrated $N-$state spin model, inspired by bent-core liquid crystals in which the frustrated behavior is well understood \cite{NE18,ME22}. The basic components of the model, located on the vertices of a triangular lattice, are Potts-like spins that can be oriented along $N$ equally-spaced orientations in the plane. The interactions of the spins are set to favor slight misalignment according to the following Hamiltonian:
\begin{equation}
\mathcal{H}=\frac{1}{2}\sum_{\text{facets}}K_1\Psi_1^2+K_3(\Psi_2-b_0)^2.\label{Hamiltonian}
\end{equation}
where $b_0=const\ne0$, which has the dimensions of inverse length \footnote{compared with the length of an edge in the lattice, $l$.}, is responsible for the frustration in the system, and the auxiliary variables $\Psi_i$ can be computed directly from the spins in each facet according to:
\begin{equation} \label{eq:4}
\begin{split}
\Psi_1=&-\frac{\theta_2-\theta_1}{l}\sin(\bar{\theta})\pm
\frac{2\theta_3-\theta_2-\theta_1}{\sqrt{3}l}\cos(\bar{\theta}),\\
\Psi_2=&\frac{\theta_2-\theta_1}{l}\cos(\bar{\theta})\pm
\frac{2\theta_3-\theta_2-\theta_1}{\sqrt{3}l}\sin(\bar{\theta}),
\end{split}
\end{equation}
where $\bar{\theta}=\left( \theta_1+\theta_2+\theta_3 \right)/3$ and the $+$ $(-)$ sign corresponds to upright (up-side-down) triangles, see Fig. \ref{fig:model}(a).

In the limit $N\to\infty$ this model recovers the recently introduced frustrated XY-spin model \cite{ME22} exhibiting cumulative frustration. 
In this limit the spin texture is interpreted as a director field and $\Psi_1$ and $\Psi_2$ are identified with the director's splay and bend fields respectively, and $b_0$ may be interpreted as the reference bend value. For finite $N$ values $\Psi_1$ and $\Psi_2$ may only assume a finite collection of states, which is further reduced by the discrete compatibility conditions. For more details regarding the manner in which this model is related to planar bent-core liquid-crystals and regarding the response in the case of the frustrated XY-spin model see \cite{ME22}. 

In order to probe the effects of discrete response modes on the frustrated conformations of minimal energy, the ground-state of the system was approached numerically. Due to the highly coupled nature of the problem the stochastic sampling of individual spins proved inefficient, and at each simulation step a pair of edge-sharing spins were sampled. The numerical process included two rounds each comprised of a simulated annealing stage combined with a global step that probed the energetic benefit associated with the rigid rotation of domains of finite size in the system. While we cannot preclude the possibility that the obtained states reflect a local minimum of the energy rather than the (possibly degenerate) true ground state, these were obtained as the configurations of minimal energy starting from multiple initial conditions, and show relative continuity as both the granularity and the number of particles are varied. For more details see supplementary material. The chosen parameters are $K_1=K_3=2$, edge length in the lattice of $l=2$ and preferred misalignment of $b_0=0.015$. Free boundary conditions were used throughout the optimization process.

The spins of a given triangle in the lattice are ordered according to panel (a) of Fig. \ref{fig:model} and their states are expressed as $[m_1,m_2,m_3]$, where $m_i$ are positive integers between $1$ and $N$ corresponding to their orientations. Note that throughout this work the local orientations were bounded away from both $0$ and $2\pi$, thus the periodicity of the angular variable did not come into play. In order to focus on the spatial gradients in the orientation we identify the minimal orientation value in each facet, $\min(m_i)=m$, and express the facet orientation as $[m+a,m+b,m+c]$ with $a,b$ and $c$ non-negative integers, at least one of which is zero. Each type of spatial gradient about $m$ corresponds to a specific triplet $(a,b,c)$ and is color-coded consistently throughout this manuscript. 

We first come to examine the threshold granularity $N=N_t$ below which it is unfavorable to introduce spatial gradients to the ground state texture.
Even for a single triangular facet, and the misalignment $b_0$, it is not always favorable to form any gradient of the orientations. The state $[m,m,m]$ represents the trivial constant solution, and the state $[m,m,m+1]$ represents the smallest possible deviation from it (along with its permutations). Given a certain value of $b_0$, it might be unfavorable to form any type of non-trivial solution as the lowest possible deviation of the orientations is too large. For the above mentioned parameters, the energetic cost of the trivial solution is $e=b_0^2$. The energetic cost of the smallest deviation from it reads:
\begin{equation} \label{eq:dev}
e=b_0^2+\frac{4 \pi^2}{3 N^2}-\frac{4 \pi b_0}{\sqrt{3} N} \sin[(m+\frac{1}{3})\frac{2 \pi}{N}].
\end{equation}
This energy is favorable whenever:
\begin{equation} 
N>\frac{\pi}{\sqrt{3} b_0 \sin[(m+\frac{1}{3})\frac{2 \pi}{N}]} > \frac{\pi}{\sqrt{3}b_0}. 
\end{equation}
Setting $b_0=0.015$ we obtain: $N_{t}>120.91$. 
Note that any system whose diameter, $D$, includes many individual particles, i.e.  $1\ll D$, yet is smaller than the geometric length-scale associated with the frustration, $D<1/b_0$, will require $b_0$ to be small and consequently $1\ll N_t$. We return to this point in the discussion.

Panel (c) of Fig. \ref{fig:model} shows the optimal energy of a single facet vs. $N$ for several gradient types. One can see that the threshold at which the trivial solution, marked with a dashed line, ceases being the most favorable reads $N_t=121$. Incorporating a deviation from a given orientation ,$m$ , may exclude repeating the same gradient type with the same $m$ value in its vicinity. As the energy depends on the value of $m$ the energetic cost of a uniform gradient in the system may vary spatially. The minimized conformations in the case of a marginal granularity of $N=124$ is shown in panel (d). The uniform gradient causes the value of $m$ to vary spatially leading the gradient orientation to deviate  away from the local average orientation and thus contribute to $\Psi_1$ rather than $\Psi_2$. By examining Eq.\eqref{eq:dev} we obtain that for $N=124$ only nine distinct values of $m$ keep the deviation energy of the gradient $[m,m,m+1]$ smaller than the energy of the uniform phase $b_0^2$. As this gradient causes the value of $m$ in adjacent lines to increase by one we conclude that 
large domains will only show nine lines of gradients, with the remainder of the domain showing the trivial solution. This result is general for  marginal values of $N_t \lesssim N$, where the incorporation of gradients is favorable only in small domains up to a finite size. Further increasing the size of the system will incorporate only domains of trivial solution rendering the energetic signature of the non-trivial regions less distinguishable, as can be seen in panel (e).

\begin{figure}
\includegraphics[width=8.6cm]{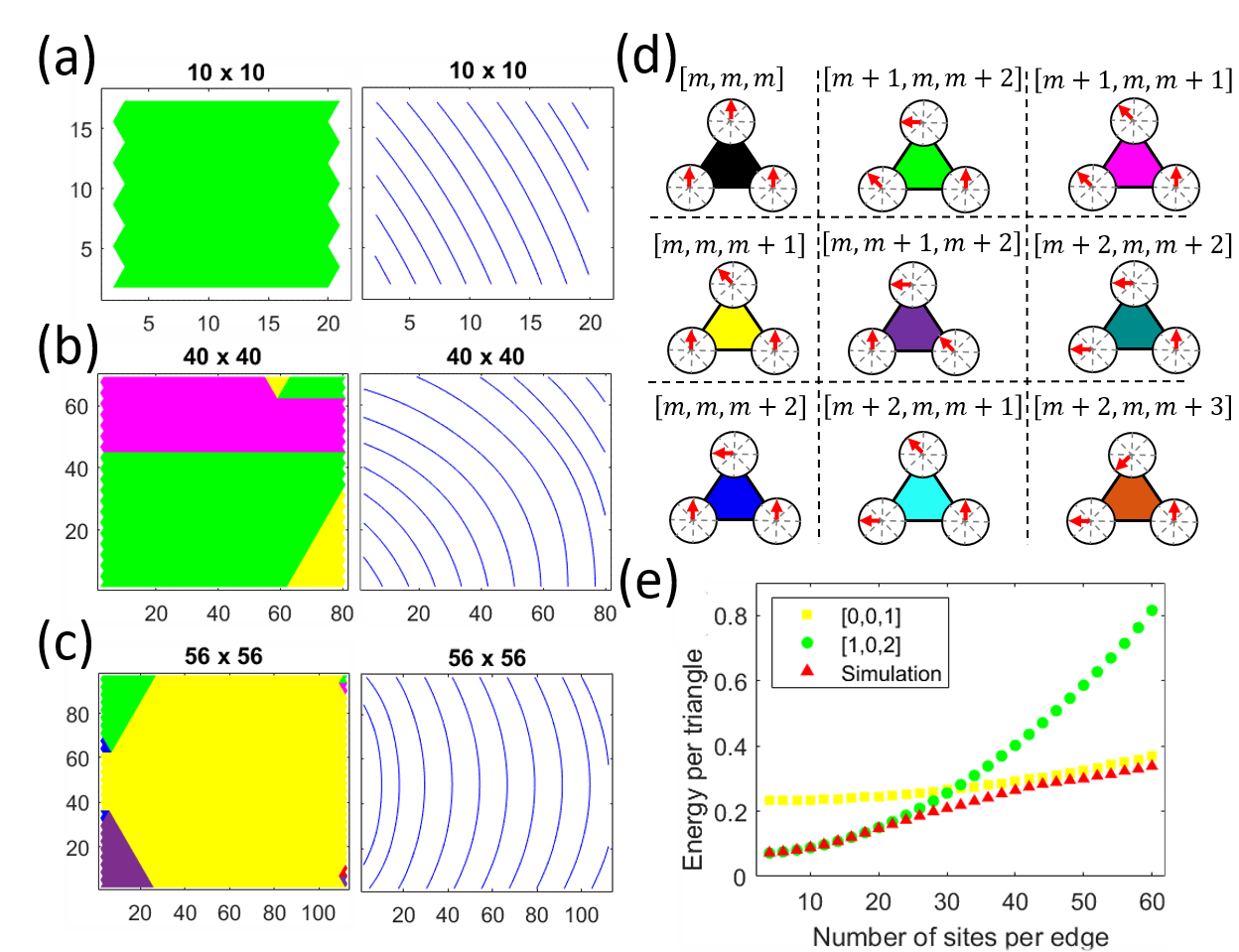}
\caption{Results for $N=356$. (a)-(c) Resulting minimal conformations for lattices of $10\times10$, $40\times40$ and $56\times56$ sites per edge, respectively (note that as stated above $l=2$). Color coding of the gradient types in the panels to the left is according to (d) and the right panels display the interpolated streamlines. (e) Energy per triangle vs. the number of sites per edge in isotropically growing lattices. The resulting conformations from the simulation marked in red triangles are shown together with the minimal energies per triangle in lattices of a single conformation where $[m,m,m+1]$ and $[m+1,m,m+2]$ are marked in yellow squares and green circles, respectively.}
\label{fig:356}
\end{figure}

As the granularity of the orientation of the spins is reduced, by increasing $N$, more complex response modes are incorporated into the ground-state. While marginal values of $N$ only display uniform gradient ground-states, i.e. the entire solution is comprised of a single gradient type $[m+a,m+b,m+c]$, for higher values of $N$ non-uniform responses appear.

Figure \ref{fig:356} displays representative samples of the ground-state solutions for $N=356$ depicting the transition between gradient types as the domain size is increased. 
From panel (c) of figure \ref{fig:model} we deduce that for $N=365$ the gradient $[m+1,m,m+2]$ is associated with the minimal energy for a single triangle, while the gradient 
$[m,m,m+1]$ is associated with the next lowest energy. 
Indeed, the ground-state of the smallest domain in panel (a) of figure \ref{fig:356} displays only a single gradient $[m+1,m,m+2]$. 
Moreover, as depicted in panel (e) of figure \ref{fig:356}, the energy associated with the uniform gradient $[m+1,m,m+2]$ increases very fast with system size, while the energy associated with the uniform gradient $[m,m,m+1]$ grows at a much slower rate. Consequently, for larger systems we expect the gradient type $[m,m,m+1]$ to dominate the ground-state solution, as seen in panel (c). Note, that the transition is not abrupt, and that the system succeeds in further lowering the elastic energy by incorporating additional gradient types.  This behaviour demonstrates the global aspects of cumulative geometric frustration, as the optimal solution for larger domains may be fundamentally different from the solutions for smaller domains.

\begin{figure}
\includegraphics[width=8.0cm]{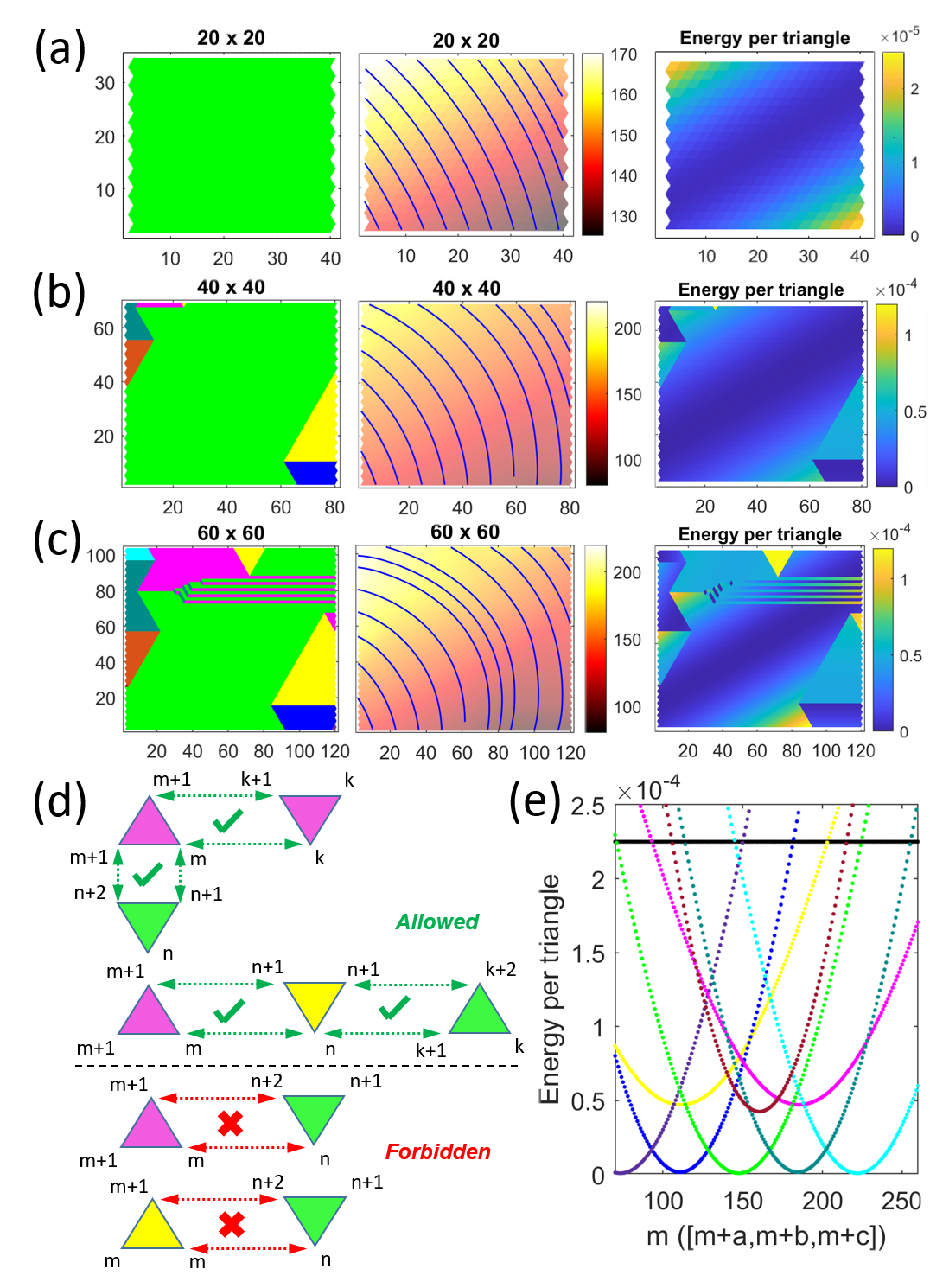}
\caption{ 
Results for $N=446$. (a)-(c) Resulting minimal conformations for lattices of (a) $20\times20$, (b) $40\times40$ and (c) $60\times60$ sites per edge. Color coding is according to panel (d) of figure \ref{fig:356}. Each triplet shows gradient types (left), interpolated streamlines along with corresponding $m$ value heat-map (middle) and energy per site (right). Notice different color-bar scales. (d) A graphic representation of the orientation dependant allowed and forbidden transitions between gradient types. (e) Energy per a single triangle of a conformation $[m+a,m+b,m+c]$ vs. $m$ in that notation. The color coding is also according to panel (d) of figure \ref{fig:356}.}
\label{fig:446}
\end{figure}

As the value of $N$ is further increased large domains display more intricate ground-states. Figure \ref{fig:446} displays the ground-state for three domain sizes for $N=446$. Panel (a) shows a relatively small domain in which the ground-state is given by the uniform gradient $[m+1,m,m+2]$, as predicted in Figure \ref{fig:356} (c) for a single triangle. Note, however, that while the gradient is uniform the energy is not; the local energy also depends on the relative orientation between the direction of the gradient and the average local orientation given by $m$. Panel (e) shows the dependence of the energy of a single triangular facet of nine gradient types on the local value of $m$. The energetically favorable gradient type for a certain $m$ value may be succeeded by another gradient type as $m$ is varied. 
In the constant gradient solution displayed in Figure \ref{fig:446} (a)
the optimal $m$ value is obtained along one of the diagonals of the domain. Away from this diagonal $m$ varies, leading to the strain energy buildup in the top-left and bottom-right corners of the domain. Figure \ref{fig:446} (b) show that as the domain size is increased the growing strain is screened by incorporating different gradient types in these energetically unfavorable regions, namely $[m,m,m+2]$ and $[m+2,m,m+2]$. Note, however, that the discrete compatibility conditions preclude the gradient type $[m,m,m+2]$ at the bottom right from having a horizontal or upper-right diagonal interfaces with the underlying  $[m+1,m,m+2]$ gradient. Consequently, it must be buffered by an additional gradient type, $[m,m,m+1]$, that bridges between the incompatible gradients. While the buffering gradient $[m,m,m+1]$ is less energetically favorable, its presence is necessitated by the discrete compatibility conditions. In the upper-left corner of Figure \ref{fig:446} (b) we similarly observe the incorporation of a buffering gradient $[m+2,m,m+3]$ to bridge between two incompatible gradients.

Panel (c) in Figure \ref{fig:446} depicts a further increase in domain size and the appearance of more intricate texture. A new gradient type $[m+1,m,m+1]$ ( marked by the pink domain) is incorporated near the top-left portion of the domain to reduce the local stain. This gradient type, while allowing horizontal interfaces with the underlying $[m+1,m,m+2]$ gradient (marked in green), cannot share any diagonal interface with it. Consequently here too a buffering gradient is required. Growing the buffering gradient $[m,m,m+1]$  (marked in yellow) along the entire incompatible interface is prohibitively energetically costly. As a result the bottom region of the (pink) gradient $[m+1,m,m+1]$ must extend all the way to the boundary of the domain. This gives rise to the (pink) striped pattern in the upper-half of panel (c) in Figure \ref{fig:446}. These stripes, while energetically unfavorable compared to the background gradient, are required by the compatibility conditions to reach a boundary.

Frustrated systems with continuous DOF are characterized by their compatibility conditions restricting the realizable local structures \cite{ME21}. In systems with discrete degrees of freedom these conditions are replaced by local matching rules between the states of adjacent sites. For a continuous two dimensional unit vector field, the compatibility condition consists of a single partial differential equation relating the local bend and splay and their oriented first derivatives \cite{NE18}. For the frustrated XY model on a triangular lattice the compatibility condition restricts the possible splay and bend values in any three side-sharing triangular facets \cite{ME22}. In the present case, as our basic building blocks are oriented gradients we encounter a nearest neighbor compatibility condition in the form of a local restrictions rule prescribing
the gradients that are allowed to share a (specific) edge with a given gradient as depicted in Fig \ref{fig:446}(d). In addition, the discrete analog of the continuous compatibility condition predicts uniquely the gradient of a facet given the gradients in two of its side sharing triangular facets.

For moderate $N$ values, where the state space of the system is limited, the system may seek to incorporate mismatching states next to one another. In such cases, energetically unfavorable buffering domains (as depicted in Fig \ref{fig:446}(b)) will form. As the energetic cost of these buffering domains becomes exceedingly large the mismatched domains will be topologically required to span the entire system to the free boundary, as in Fig \ref{fig:446}(c).      
 
\begin{figure}
\includegraphics[width=7.6cm]{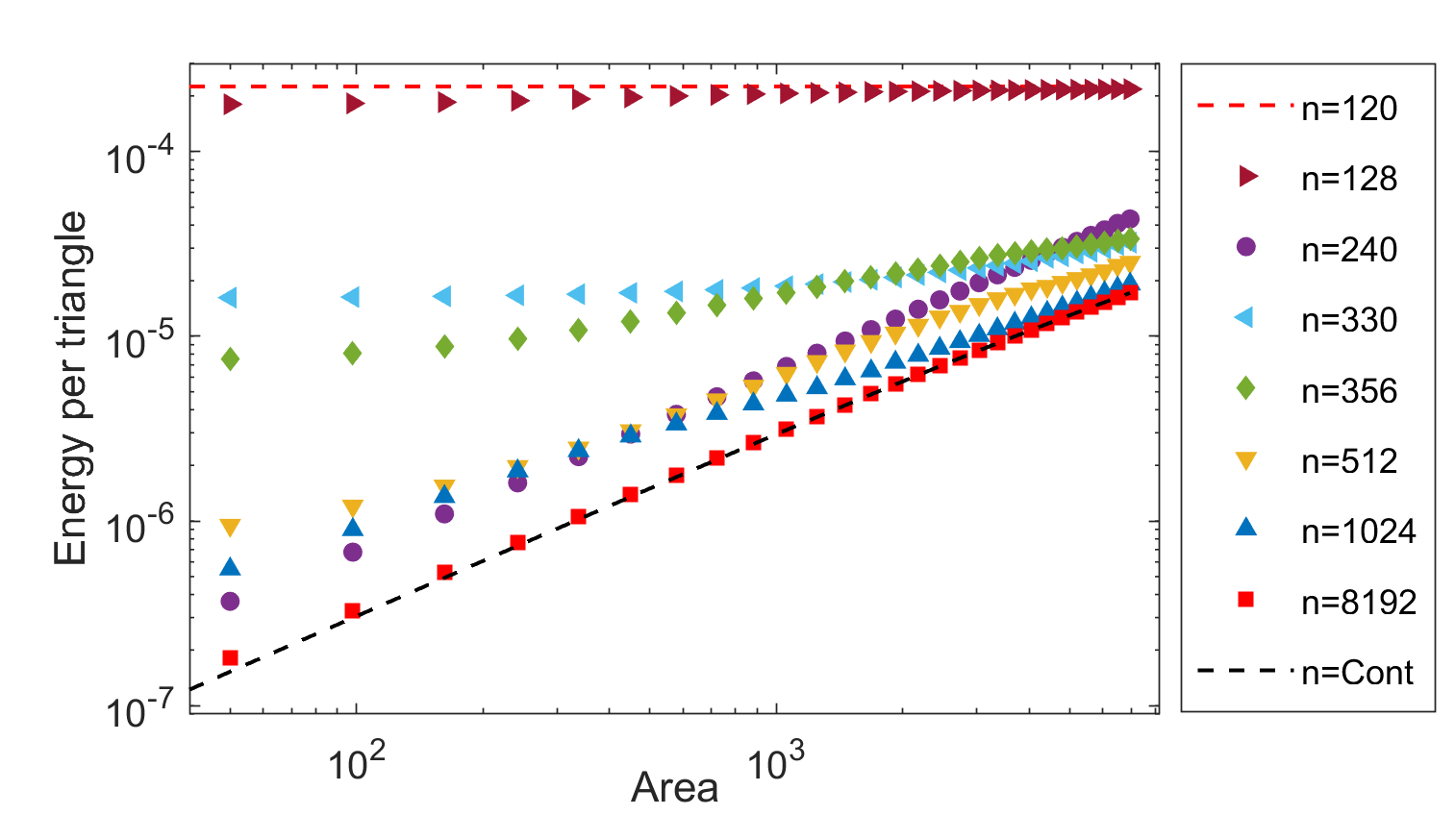}
\caption{ Logarithmic plot of the energy per triangle of the ground-states resulting from the simulation vs. the area of isotropic domains for different directions granularity values, $N$. The black dashed line (bottom) shows the results in the case of the continuous XY-spins, as was presented in \cite{ME22}. }
\label{fig:4}
\end{figure}

Considering the energy scaling with system size we  observe the expected transition from a uniform resolution of the frustration at low $N$ values to a highly cooperative ground-states exhibiting a super-extensive energy scaling at large $N$ values. Large enough $N$ values recover quantitatively the behavior of the corresponding continuous frustrated XY-spin system where the energy per particle scales as: $E/A\propto A$, where $A$ is the area of the system \cite{ME22}. Figure \ref{fig:4} shows that for $N=8192$ the system follows closely the super-extensive energy scaling of the continuous case, while the system at $N=120$ follows an extensive energy scaling. The transition between these two limiting behaviors is, however, non-monotonic. For example, while $N=128$ follows closely the extensive scaling, $N=240$ displays a super-extensive scaling quite similar to the continuous case, and $N=330$ again shows only small deviations from an extensive behavior. Na\"ively, one would expect that as $N$ is increased, more refined gradients could be constructed to better approximate the continuous case; this is indeed the case for large $N$ values. For moderate $N$-values, however, ``resonant"-like values of $N$ could approximate
the attempted gradient $b_0$ almost perfectly, while nearby increased values such as $N+1$ will not perform as well. These discrete effects, most prominently manifest for moderate $N$-values also affect the compatibility conditions and the rate at which the energy of a given gradient varies as $m$ is varied. As the resonant mechanism responsible for the non-monotonous approach to the continuum limit is general, we expect  
this phenomenon to appear in other frustrated systems at moderate levels of granularity.

The frustrated N-state spin model presented here allows us to bridge the gap between the cumulative frustration of the frustrated XY-model, and the non-cumulative frustration of the Ising anti-ferromagnet on a triangular lattice.
Up to $N_t=121$ all the spins in the ground-state are co-aligned, and the ground-state energy is extensive. The threshold value $N_t$ is inversely proportional to the relatively small strength of the frustration $b_0 l\ll 1$. While increasing $b_0$ may lower $N_{t}$, it also decreases the length-scale associated with the geometric frustration and thus diminishes the domain size at which frustration saturation is met. Eventually, when the geometric length-scale associated with the frustration becomes shorter than a single unit cell, no characteristics of the cumulative frustration will remain \cite{ME21}. 
 
For large values of $N$ the cumulative frustration behavior of the continuum limit is recovered and the granularity of the DOF does not play a significant role in shaping the ground-state. However, for moderate values of $N$, while we observe cumulative frustration and a highly cooperative ground-state, the discrete nature of the DOF leads to qualitatively different ground-states and exotic response properties that are not observed in the continuous case. These include the formation of striped patterns and the
appearance of buffering moderate energy regions that facilitate the incorporation of low energy domains whose incorporation is otherwise precluded by the discrete matching rules. These motifs are unique to the  discrete-frustrated-spin system and have no analogue in the continuous frustrated XY-model.

\begin{acknowledgments}

\end{acknowledgments}

\bibliographystyle{unsrt}
\bibliography{main.bib}

\end{document}